\documentclass{article}
\usepackage{amsmath,psfig,amssymb}
\newcommand{\ol}{\setlength{\itemsep}{0pt.}\begin{enumerate}}
\newcommand{\eol}{\end{enumerate}\setlength{\itemsep}{-\parsep}}
\newcommand{\remove}[1]{}

\newcommand{\ber}{{\begin{eqnarray*}}}
\newcommand{\eer}{{\end{eqnarray*}}}

\newcommand\nc\newcommand
\nc\bfa{{\bf a}}\nc\bfA{{\bf A}}\nc\cA{{\cal A}}
\nc\bfb{{\bf b}}\nc\bfB{{\bf B}}\nc\cB{{\cal B}}
\nc\bfc{{\bf c}}\nc\bfC{{\bf C}}\nc\cC{{\cal C}}
\nc\bfd{{\bf d}}\nc\bfD{{\bf D}}\nc\cD{{\cal D}}
\nc\bfe{{\bf e}}\nc\bfE{{\bf E}}\nc\cE{{\cal E}}
\nc\bff{{\bf f}}\nc\bfF{{\bf F}}\nc\cF{{\cal F}}
\nc\bfg{{\bf g}}\nc\bfG{{\bf G}}\nc\cG{{\cal G}}
\nc\bfh{{\bf h}}\nc\bfH{{\bf H}}\nc\cH{{\cal H}}
\nc\bfi{{\bf i}}\nc\bfI{{\bf I}}\nc\cI{{\cal I}}
\nc\bfj{{\bf j}}\nc\bfJ{{\bf J}}\nc\cJ{{\cal J}}
\nc\bfk{{\bf k}}\nc\bfK{{\bf K}}\nc\cK{{\cal K}}
\nc\bfl{{\bf l}}\nc\bfL{{\bf L}}\nc\cL{{\cal L}}
\nc\bfm{{\bf m}}\nc\bfM{{\bf M}}\nc\cM{{\cal M}}
\nc\bfn{{\bf n}}\nc\bfN{{\bf N}}\nc\cN{{\cal N}}
\nc\bfo{{\bf o}}\nc\bfO{{\bf O}}\nc\cO{{\cal O}}
\nc\bfp{{\bf p}}\nc\bfP{{\bf P}}\nc\cP{{\cal P}}
\nc\bfq{{\bf q}}\nc\bfQ{{\bf Q}}\nc\cQ{{\cal Q}}
\nc\bfr{{\bf r}}\nc\bfR{{\bf R}}\nc\cR{{\cal R}}
\nc\bfs{{\bf s}}\nc\bfS{{\bf S}}\nc\cS{{\cal S}}
\nc\bft{{\bf t}}\nc\bfT{{\bf T}}\nc\cT{{\cal T}}
\nc\bfu{{\bf u}}\nc\bfU{{\bf U}}\nc\cU{{\cal U}}
\nc\bfv{{\bf v}}\nc\bfV{{\bf V}}\nc\cV{{\cal V}}
\nc\bfw{{\bf w}}\nc\bfW{{\bf W}}\nc\cW{{\cal W}}
\nc\bfx{{\bf x}}\nc\bfX{{\bf Z}}\nc\cX{{\cal X}}
\nc\bfy{{\bf y}}\nc\bfY{{\bf Y}}\nc\cY{{\cal Y}}
\nc\bfz{{\bf z}}\nc\bfZ{{\bf Z}}\nc\cZ{{\cal Z}}

\def\wt{\qopname\relax{no}{wt}}
\def\tr{\qopname\relax{no}{tr}}
\def\Tr{\qopname\relax{no}{Tr}}
\def\dim{\qopname\relax{no}{dim}}

\begin{document} 
\twocolumn

\title       { Nonbinary Quantum Stabilizer Codes }

\author      { Alexei Ashikhmin
\thanks {Bell Laboratories, Lucent Technologies, 
  600 Mountain Ave., Rm.: 2C-180,
  Murray Hill, NJ 07974. } \and 
 %{ E-mail: aea@research.bell-labs.com} 
Emanuel  Knill
\thanks { Los Alamos National Laboratory 
  Group CIC-3, Mail Stop B265, 
Los Alamos, NM 87545. 
%{E-mail: knill@c3serve.c3.lanl.gov}
}
}
\date{}
\maketitle
\begin{abstract}
We define and show how to construct nonbinary quantum stabilizer
codes. Our approach is based on nonbinary error bases. It
generalizes the relationship between selforthogonal codes over $\bfF_{4}$
and binary quantum codes to one between selforthogonal codes over
$\bfF_{q^2}$ and $q$-ary quantum codes for any prime power $q$.
\end{abstract}
 

\section{Introduction}

Probably the most important class of binary quantum codes are quantum
stabilizer codes. They play a role similar to the linear codes in
classical coding theory.  Quantum stabilizer codes have simple
encoding algorithms, can be analyzed using classical coding theory,
and yield methods for fault tolerant quantum computation.
The first examples of quantum codes found by Shor~\cite{shor95},
and Steane~\cite{stean96_1,stean96_2} were quantum stabilizer
codes. General quantum stabilizer codes were introduced by
Gottesman~\cite{Gottesman96} and Calderbank
\emph{et. al.}~\cite{Calderbank97}. Later Calderbank
\emph{et. al.}~\cite{Calderbank98} gave the now standard connection
between quantum stabilizer codes and classical selforthogonal codes,
which was used to construct a number of new good quantum codes.

While the theory of binary quantum stabilizer codes is now well
developed, nonbinary codes have been relatively ignored.  A connection
between classical codes over $\bfZ_n$ and quantum codes is given
in~\cite{Knill96_1,Knill96_2}. The connection is based on a
stabilizer construction derived from so-called nice error
bases. Raines~\cite{rains} obtained a number of results for $p$-ary
($p$ prime) quantum stabilizer codes generalizing the $\bfF_{4}$
constructions for binary quantum codes.

Here we consider the problem of constructing $p^m$-ary quantum codes
from classical selforthogonal codes over $\bfF_{p^{2m}}$. The notion
of selforthogonality arises naturally from the error bases
of~\cite{Knill96_1,Knill96_2} and can be identified with that arising
from a field-theoretically defined simplectic form. Good
selforthogonal codes with respect to this form have already been found
by Bierbrauer and Edel~\cite{Bierbrauer98}, and our construction can
be used to obtain associated quantum codes.

\section{Basic Definitions}

We start with the basic notions of classical and quantum coding
theory. Denote by $\bfF_{p^m}$ the Galois field of $p^m$ elements,
where $p$ is a prime number and $m$ is an integer. Let $\alpha_1,
\alpha_1, \ldots , \alpha_m$ denote the elements of a basis of
$\bfF_{p^m}$ over $\bfF_{p}$.  We fix
a non-zero $\bfF_{p}$-linear functional
$\tr: \bfF_{p^m}\rightarrow \bfF_{p}$
(called a \emph{trace function}). Thus $\tr$ satisfies
\begin{eqnarray*}
tr(a+b)&=&tr(a)+tr(b),\\
tr(\alpha a)&=&\alpha tr(a),
\end{eqnarray*}
for all $a,b\in \bfF_{p^m}, \alpha \in \bfF_p$.
Note that for $x\in\bfF_{p^m}$, $\tr_x(a) = \tr(xa)$
defines another trace function, and that all such functions
can be obtained this way. The standard trace function
is the one defined by viewing $\bfF_{p^m}$ as an extension
of $\bfF_p$ and letting $\tr(a)=\sum_{i=0}^{m-1}a^{p^i}$,
\cite[Chapter 2.3]{macw77}.

Let $t$ divide $m$.  A classical $p^t$-linear code $C$ over a field
$\bfF_{p^m}$ of length $n$ and size $(p^t)^k$, is a $k$ dimensional
$p^t$-linear subspace of the space $\bfF_{p^m}^n$. In other words, for
any $\bfa,\bfb$ from $C$ and any $\alpha,\beta\in \bfF_{p^t}$ the
vector $\alpha \bfa+\beta \bfb$ is also from $C$.  Let $*$ be a
$\bfF_{p^t}$-bilinear form (an \emph{inner product}).  A code $C$ is
selforthogonal for $*$ if for all vectors $\bfa$ and $\bfb$ from $C$
the following property holds
\begin{equation}\label{eq:inpr0}
\bfa * \bfb=0.
\end{equation}
The code $C^\perp=\{\bfv: \bfv*\bfa=0 \mbox{ for } \forall \bfa\in C\}$ is
called dual of $C$ with respect to (\ref{eq:inpr0}).

{\bf Remark} For an introduction to the theory of Galois Fields and
classical codes see e.g. \cite{macw77}.

A $q$-ary quantum code $Q$ of length $n$ and size $K$ is a
$K$-dimensional subspace of a $q^n$-dimensional Hilbert space.  This
Hilbert space is identified with the $n$-fold tensor product of
$q$-dimensional Hilbert spaces. The $q$-dimensional spaces are thought
of as the state spaces of \emph{$q$-ary systems} in the same way as
the values $0$ and $1$ can be thought of as the possible states of a
bit in a bit string.  We identify the state spaces with the
$q$-dimensional complex linear space $\bfC_q$.  An important
characteristic of a quantum code is its minimum distance. If a code
has minimum distance $d$ then it can detect any $d-1$ and correct any
$\lfloor \frac{d-1}{2} \rfloor$ errors. As a result it is desirable to
keep $d$ as large as possible. A strict definition of the minimum
distance is given in the next section after introducing error bases.

{\bf Remark} For introductions to the theory of quantum error
correcting codes see e.g. \cite{knill97,Gottesman97,knill99}.
For a reader with a background in classical coding theory 
the papers \cite{Ashikhmin99,Ashikhmin00_1,Ashikhmin00_2} have
brief introductions to the field. 

\section{Error Basis}\label{error_basis}

A general quantum error of a $p^m$-ary quantum system, is a linear
operator acting on the space $\bfC_{p^m}$. If $|\bfv\rangle$ is a state (a
unit vector in the space) of the system, then the effect of error $E$
is to transform it to the state $E |\bfv\rangle$.  It is convenient to
confine ourselves to errors that form a basis of the vector space of
linear operators acting on $\bfC_{p^m}$. Let linear operators $e_1,
e_2, \ldots, e_{p^{2m}}$ form such a basis. 
If $|\bfv\rangle$ represents a
state of $n$ $p^m$-ary systems it can be altered by an error operator of
the form
\begin{equation}\label{E}
E=\sigma_1\otimes \sigma_2\otimes \ldots \otimes \sigma_n,
\end{equation}
where $\sigma_i\in \{e_1, e_2, \ldots, e_{p^{2m}} \}$.  A
general error operator is a linear operator acting on the $n$-fold
tensor product of $\bfC_{p^m}$. Any such operator
can be written down as a linear combination of error operators of the
form (\ref{E}).  It is well known from the general theory of quantum
codes that if a code can correct a given set ${\cal E}$ of
error operators, then it can correct the linear span of ${\cal E}$.
For this reason it makes sense to focus on operators of the form
(\ref{E}).

It is always possible to determine operators $e_1,$ $e_2, \ldots,$
$e_{p^{2m}}$ in such a way that one of them, say $e_1$, is the
identity operator $I_{p^m}$. Define the {\em weight} of $E$ in
(\ref{E}) as
\begin{equation}\label{wt}
\wt(E)=|\{ \sigma_i\not =I_{p^m}\}|.
\end{equation}
In the depolarizing channel model of errors~\cite{bennett96}, the
operators $e_2,e_3,\ldots$ satisfy $\Tr(e_i^\dagger e_j) =
p^m\delta_{i,j}$, where $\Tr$ is the trace of linear operators. When
transmitting a qubit through a depolarizing channel, the probability
that it is untouched (i.e. affected by the identity operator) is $1-r$
and the probability that it is affected by $e_i,i>1$, is
$r/(p^{2m}-1)$.  Thus, the probability of an error operator decreases
exponentially with weight, a feature common to most realistic error
models~\cite{knill99}. This explains why it is desirable to correct or
detect all error operators up to some given weight.

Let $P$ be the orthogonal projection operator onto $Q$.  It can be
shown that (see e.g.~\cite{Knill96_1}) an error operator $E$ is
detectable by $Q$ iff
\begin{equation}\label{eq:err}
PEP=c_E P. 
\end{equation}
The largest integer $d$ such that every error of weight $d-1$ or less
can be detected by a code is called its minimum distance.

We now define an explicit error basis for $p^m$-ary quantum codes. Let
$T$ and $R$ be linear operators acting on the space $\bfC_{p}$ defined
by the matrices with entries
$$
T_{i,j}=\delta_{i,j-1 \mbox{\small mod }p} \mbox{ and } R_{i,j}=\xi^i\delta_{i,j},
$$
where $\xi=e^{ \iota  2\pi/ p}, \iota=\sqrt{-1}$ and the indices
range from $0$ to $p-1$~\cite{Knill96_1}. 
It is easy to check that
$$
TR=\xi RT
$$
and therefore 
\begin{eqnarray}
T^iR^j&=&\xi^{ij}R^jT^i,\\
\label{eq:pro}
\left(T^iR^j\right)\left(T^kR^l\right)
  &=&\xi^{il-jk}\left(T^kR^l\right)\left(T^iR^j \right),\\
\label{eq:grpr}
\left(T^iR^j\right)\left(T^kR^l\right)&=&\xi^{-jk} T^{i+k}R^{j+l}.
\end{eqnarray}
The Hermitian transposes of $T^i$ and $R^i$ are obtained by raising to the
power $p-1$:
\begin{equation}\label{eq:dag}
(T^i)^\dag =(T^i)^{p-1}, (R^i)^\dag =(R^i)^{p-1}.
\end{equation}
Note that 
\begin{equation}\label{eq:Ip}
T^p=R^p=I_p.
\end{equation}
From (\ref{eq:grpr}) and (\ref{eq:Ip}) it follows that for $p>2$ 
\begin{equation}\label{eq:p_power}
(T^iR^j)^p=\xi^{-ij(1+2+\ldots +(p-1))}=I_p.
\end{equation}

Since $\Tr(T^iR^j)=0$ except when $i=j=0\mod p$, the operators
$T^iR^j$ form an orthogonal operator basis under the usual inner
product for operators given by $\langle A,B\rangle = \Tr(A^\dagger B)$.
Let $a,b\in \bfF_{p^m}$. Using a basis of $\bfF_{p^m}$ over
$\bfF_{p}$, we can write uniquely
\begin{eqnarray*}
a& = &a_1\alpha_1+a_2\alpha_2+\ldots +a_m\alpha_m, \\
b& = &b_1\alpha_1+b_2\alpha_2+\ldots +b_m\alpha_m,
\end{eqnarray*}
with the $a_i$ and $b_i$ in $\bfF_p$. Define
$$
T_aR_b=(T^{a_1}\otimes T^{a_2} \otimes\ldots    \otimes T^{a_m})
 (R^{b_1}\otimes R^{b_2}  \otimes\ldots    \otimes R^{b_m}).
$$
The operators $T_aR_b$ then form an orthonormal basis. 
The multiplication rules given above can be generalized.
Define 
\begin{equation}\label{eq:inpr1}
\langle a,b\rangle=\sum_{i=1}^m a_ib_i \in \bfF_p.
\end{equation}
From (\ref{eq:grpr}) and the identity $(A\otimes
B)(C\otimes D)=AC\otimes BD$ it follows that 
\begin{equation}\label{eq:grpr:ten}
 (T_aR_b)(T_cR_d)=\xi^{-\langle b,c\rangle}T_{a+c}R_{b+d}.
\end{equation} 
(\ref{eq:pro}) and (\ref{eq:inpr1}) yield
\begin{equation}\label{eq:prod:ten}
 (T_aR_b)(T_cR_d)=\xi^{\langle a,d\rangle-\langle b,c\rangle}(T_cR_d)(T_aR_b). 
\end{equation}

\section{Nonbinary Stabilizer Codes}

Let $\bfa^\dag=(a^{(1)},a^{(2)},\ldots ,a^{(n)})
,\bfb^\dag=(b^{(1)},b^{(2)},\ldots ,b^{(n)})$ be vectors from the
space $\bfF_{p^m}^n$. (Throughout this section we use superscripts to
label the systems.)  As discussed in the previous section, it is
enough to consider the error operators given by
\begin{equation}\label{eq:E:ten}
E_{\bfa,\bfb}=T_{a^{(1)}}R_{b^{(1)}}\otimes T_{a^{(2)}}R_{b^{(2)}}\otimes \ldots
\otimes T_{a^{(n)}}R_{b^{(n)}}.
\end{equation} 
The set of operators ${\cal E} = \{\xi^i E_{\bfa,\bfb}|0\le i\le p-1\}$
form a group of order $p^{2mn+1}$. The center ${\cal Z}$  of ${\cal E}$ is 
generated by $\xi I$ and therefore has order $p$.
For vectors $\bfa, \bfd \in
\bfF_{p^m}^{n}$ define an inner product by
\begin{equation}\label{eq:inpr2}
\langle \bfa, \bfd\rangle=\sum_{i=1}^n \langle a^{(i)},d^{(i)}\rangle,
\end{equation}
where $\langle a^{(i)},d^{(i)}\rangle$ is defined in (\ref{eq:inpr1}). 
It follows from (\ref{eq:prod:ten}) that  
\begin{equation}\label{eq:trans:E}
E_{\bfa,\bfb} E_{\bfc,\bfd}= \xi^{  \langle \bfa, \bfd\rangle-\langle \bfb,\bfc\rangle}
E_{\bfc,\bfd}E_{\bfa,\bfb}.
\end{equation}
From (\ref{eq:grpr:ten}) we have 
\begin{equation}\label{eq:prod:E}
E_{\bfa,\bfb} E_{\bfc,\bfd}= \xi^{ -\langle \bfb,\bfc\rangle}
 E_{\bfa+\bfc, \bfb+\bfd}.
\end{equation}
From (\ref{eq:E:ten}) and (\ref{eq:p_power}) it follows that for any 
$\bfa$ and $\bfb$ and $p>2$,
\begin{equation}\label{eq:E:p_power}
E_{\bfa,\bfb}^p=I_{p^{mn}}.
\end{equation}

Quantum {\em stabilizer codes\/} are defined as joint eigenspaces of
the operators of a commutative subgroup $S$ of ${\cal E}$.  Without
loss of generality, assume that ${\cal Z}\subseteq S$.  If this is not
the case, extend $S$ by ${\cal Z}$.  The order of $S$ is a power of
$p$, $|S|=p^{r+1}$. The joint eigenspaces of $S$ are
associated with linear characters $\mu$ of the group $S$ whose value
$\mu(E)$ is the eigenspace's eigenvalue with respect to $E$. Clearly
it must be the case that $\mu(\xi I)=\xi$.  Let $\mu$ be any one of
the $p^r$ characters of $S$ which satisfy this constraint.
We define a quantum stabilizer code $Q$ as the eigenspace 
associated with $\mu$. To determine the dimension of
$Q$, consider the orthogonal
projection  operator $P$ on $Q$, which can be written  in the form
$$
P=\frac{1}{|S|}\sum_{E\in S}
\bar\mu(E) E. 
$$
Since for $E\in {\cal E}\setminus{\cal Z}$, $\Tr E = 0$, we have
\begin{eqnarray*}
\dim Q &=& \Tr P\\
       &=& \frac{1}{|S|}\sum_{i=0}^{p-1}\bar\mu(\xi^i I)\Tr(\xi^i I)\\
       &=& \frac{1}{p^{r+1}}\sum_{i=0}^{p-1}p^{mn}\\
       &=& p^{mn-r}.
\end{eqnarray*}
Hence $Q$ is an $[[n,mn-r]]_{p^m}$ quantum stabilizer code.

We next establish a connection between quantum stabilizer and
classical selforthogonal codes.  Note that since the error basis is
obtained as a tensor product of $p$-ary error bases, stabilizer codes
can be viewed as standard $p$-ary stabilizer codes. This situation is
essentially the same as for classical linear codes over $\bfF_{p^m}$.
However, since the goal is to protect against errors on $p^m$-ary
systems, we wish to usefully relate $p^m$-ary stabilizer codes to
classical codes over $\bfF_{p^{2m}}$.

First we show how to construct a classical code from a quantum code. 
Let $\varphi$ be an isomorphism of the vector space $\bfF_p^m$.
Clearly the set $C=\{(\bfa, \varphi^{-1}\bfb)|
E_{\bfa,\bfb}\in S\}$ is an $\bfF_p$-linear code of length $2n$ and
size $p^r$.  Moreover, since all operators from $S$ commute the
following property holds for any two vectors $(\bfa,\bfb)$ and
$(\bfa',\bfb')$ from $C$
\begin{equation}\label{eq:inprcode} 
\langle  \bfa, \varphi(\bfb')\rangle-\langle \bfa',
\varphi(\bfb)\rangle=0.
\end{equation}
Thus $C$ is selforthogonal with respect to the inner product defined
by $(\bfa,\bfb)*(\bfa',\bfb')=\langle \bfa,
\varphi(\bfb')\rangle-\langle \bfa', \varphi(\bfb)\rangle$. Later we
will choose $\varphi$ to relate the inner product to the structure of
$\bfF_{p^m}$.

The minimum distance of a stabilizer code defined by $S$ is
related to the classical minimum distance of $C^\perp\setminus C$, where
$C^\perp$ is the dual code of $C$ with respect to
(\ref{eq:inprcode}).
Define the weight of $\bfv=(\bfa,\bfb)\in \bfF_{p^m}^{2n}$ as
$$
\wt(\bfv )=|\{i: a^{(i)}\not =0 \mbox{ or } b^{(i)}\not =0 \}|.
$$
Using arguments similar to ones from \cite{Calderbank97}, one can show that 
the minimum distance of a stabilizer code of $S$ equals
$\min\{\wt(\bfv):\bfv \in C^\perp\setminus C\}$. For completeness
we give a general proof of this fact.

Denote by $S^\perp$ the group of operators in ${\cal E}$ that commute
with all operators from $S$. Thus $S^\perp$ is given by $S^\perp =
\{\xi^i E_{\bfa,\bfb}: (\bfa,\bfb)\in C^\perp\}$.  The desired fact
follows from the observation that $E'\in{\cal E}$ is detectable iff
$E'\not\in S^\perp\setminus S$. Let $P$ be as defined
earlier. We consider three cases.
\begin{itemize}
\item[1.] Let $E'\in S$. Then 
\begin{eqnarray}
E'P &=&  \frac{1}{|S|}\sum_{E\in S}\bar\mu(E) E'E \nonumber\\
     &=& \frac{1}{|S|}\sum_{E\in S}\bar\mu((E')^\dagger E) E\nonumber\\
     &=& \mu(E')P\label{eq:ep=xip},
\end{eqnarray}
where the last equality follows from linearity of $\mu$. 
Thus 
$$
PE'P=\mu(E')P
$$
and hence  $E'$ is detectable.

\item[2.] Let $E'\not \in S^\perp$.  Let $S_i$, $0\leq i < p$, be
defined by $S_i = \{ E\in S : E'E = \xi^i EE'\}$.  Then from
(\ref{eq:trans:E}) and the assumption, it follows that $|S_i| = |S|/p$.
Thus
\begin{eqnarray}
|S|PE'P &=& \sum_{E\in S}\bar\mu(E) EE' P\nonumber\\
     &=& E'\sum_{i=0}^{p-1}\sum_{E\in S_i} \xi^i\bar\mu(E) E P\nonumber\\
     &=& E'\sum_{i=0}^{p-1}\sum_{E\in S_i} \xi^i P\nonumber\\
     &=& E'\sum_{i=0}^{p-1}\xi^i P|S|/p\\
     &=& 0,
\end{eqnarray}
where we used (\ref{eq:ep=xip}) in the third to last step.
Again, $E'$ is detectable.

\item[3.] Let $E'\in S^\perp\setminus S$.  By taking $T$ to be the
commutative subgroup generated by $S$ and $E'$ and extending the
character $\mu$ to $T$, a subcode $Q'$ of $Q$ is obtained
corresponding to the extended character. The dimension of $Q'$ is
smaller by a factor of $p$, which implies that $Q$ is not an
eigenspace of $E'$. Since $E'$ commutes with $S$, $E'$ preserves $Q$.
All of this implies that $PE'P$ is not proportional to $P$.
\end{itemize}

The inner product defined in (\ref{eq:inprcode}) depends on the
isomorphism $\varphi$. Clearly, the set of codes obtained does
not depend on $\varphi$, so the choice of $\varphi$ is primarily
one of convenience. We now standardize this choice to simplify
the construction of large minimum distance codes.
With respect to our distinguished basis of $\bfF_{p^m}$, $\varphi$ is
given by an $m\times m$ matrix $M$ over $\bfF_p$. Choose $M$ by defining
$$
M_{i,j}=\tr(\alpha_i\alpha_j).
$$
With
$a^T=(a_1,a_2,\ldots ,a_m),b^T=(b_1,b_2,\ldots ,b_m) \in
\bfF_{p^m}$, we compute
\begin{eqnarray*}
a^T M b &=& \sum_{i=1}^m \sum_{j=1}^m
a_ib_j\tr(\alpha_i\alpha_j)\\
&=&\sum_{i=1}^m \sum_{j=1}^m 
\tr( a_ib_j \alpha_i\alpha_j )\\
&=& \tr  \left( \left(\sum_{i=1}^m a_i\alpha_i \right)  
                \left(\sum_{i=1}^m b_i\alpha_i \right)  
         \right)\\
&=&\tr(ab),
\end{eqnarray*}
where the product in the trace is multiplication in $\bfF_{p^m}$.
For vectors $\bfa$ and $\bfb$ in $\bfF_{p^m}^n$, let
$\langle \bfa,\bfb\rangle_{*}=\sum_i a^{(i)}b^{(i)}$.
With this choice of $\varphi$,
$C$ is therefore selforthogonal with respect to the inner
product defined by
\begin{equation}\label{eq:trinpr}
(\bfa,\bfb)*(\bfa',\bfb')=\tr(\langle \bfa,\bfb'\rangle_*-\langle \bfa',\bfb\rangle_*).
\end{equation}

We can now construct a quantum stabilizer code from a classical
selforthogonal code $C,|C|=p^r$. Let vectors $\bfv_i=(\bfa_i,\bfb_i),
0\le i \le r-1$ form a basis of $C$ over $\bfF_p$. Then the $p^r$
operators $E_{\bfa_i,\phi(\bfb_i )}$ together with $\xi I_{p^{mn}}$
generate a group of commuting operators of order $p^{r+1}$, which
defines $[[n,mn-r]]_{p^m}$ stabilizer codes with minimum distance
$d=\min\{\wt (\bfv):\bfv\in C^\perp\setminus C \}$.

In \cite{Bierbrauer98} a number of families of good classical codes
that are selforthogonal with respect to the inner product
\begin{equation}\label{eq:birinpr}
(\bfa,\bfb)*(\bfa',\bfb')=\langle \bfa,\bfb'\rangle_*-\langle \bfa',\bfb\rangle_*
\end{equation}
where constructed.
Since a code that is selforthogonal with respect to (\ref{eq:birinpr})
is also selforthogonal with respect to (\ref{eq:trinpr}), our results
establish a previously missing connection between the classical codes
defined in \cite{Bierbrauer98} and quantum codes.
Thus we already have many good nonbinary stabilizer
codes. For instance from \cite{Bierbrauer98} we can obtain
quantum stabilizer codes with parameters $[[q^r,q^r-(r+2),3]]_q,
~~[[q^2+1,q^2-3,3]]_q,
~~[[(q^{r+2}-1)/(q^2-1),(q^{r+2}-1)/(q^2-1)-(r+2),3]]_q \mbox{ ($r$ is 
even) },~~[[q^3(q^{r-1}-1)/(q^2-1),q^3(q^{r-1}-1)/(q^2-1)-(r+2),3]]_q
\mbox{ ($r$ is odd) }$, and others.

In conclusion, we
note that if a code is $\bfF_{p^{m}}$-linear and is
selforthogonal with respect to (\ref{eq:trinpr}) then it is
automatically selforthogonal with respect to (\ref{eq:birinpr}).
Since this does not hold for general $\bfF_p$-linear codes, one
expects to find better codes selforthogonal with respect to
(\ref{eq:trinpr}) in this class.

\noindent{\bf Acknowledgements.}
E. K. was supported by funding from NSA and DOE.


\begin{thebibliography}{99}
 

\bibitem{Ashikhmin99} A. Ashikhmin and  S. Litsyn, ``Upper bounds of the 
size of quantum codes,'' {\it IEEE Trans. Info. Theory}, vol 45, 
no. 4, pp.1205-1215, 1999.

\bibitem{Ashikhmin00_1} A. Ashikhmin, A. Barg, E. Knill,  and  S. Litsyn,
``Quantum Error Detection I: Statement of the Problem ,'' {\it IEEE
Trans. Info. Theory}, to appear .


\bibitem{Ashikhmin00_2} A. Ashikhmin, A. Barg, E. Knill,  and  S. Litsyn,
``Quantum Error Detection II: Bounds ,'' {\it IEEE
Trans. Info. Theory}, to appear .

\bibitem{bennett96} 
C. H. Bennett, D. P. DiVincenzo, J. A. Smolin and W. K. Wootters,
``Mixed state entanglement and quantum error-correcting codes,''
{\em Phys. Rev. A}, vol. 54, pp. 3824--, 1996.

\bibitem{Bierbrauer98} J. Bierbrauer and Y. Edel, ``Quantum Twisted
Codes,'' {\em preprint}, 1998. (The paper is available at
``http://www.math.mtu.edu/~jbierbra/''.)

\bibitem{Calderbank97} A.R. Calderbank, E.M. Rains, P.W. Shor and N.J.A. Sloane,  
 ``Quantum error correction and orthogonal geometry,'' 
 {\em Phys. Rev. Lett.}, vol. 78, pp. 405-409, 1997. 

 \bibitem{Calderbank98} 
 A.R. Calderbank, E.M. Rains, P.W. Shor and N.J.A. Sloane, ``Quantum errors 
correction via codes 
over $GF(4)$,''{\it IEEE Trans. Info. Theory}, vol. 44, pp.1369
--1387, 1998. 


\bibitem{Gottesman96} D. Gottesman, ``A class of quantum 
error-correcting codes saturating the quantum Hamming bound,'' 
{\em  Phys. Rev. A}, vol.54, pp. 1862-1868, 1996. 

\bibitem{Gottesman97} D. Gottesman, ``Stabilizer Codes and Quantum
Error Correction,'' {\em Ph.D. Thesis,} California Institute of
Technology,
Pasadena, California, 1997.  

\bibitem{Knill96_1} E. Knill, ``Non-binary Unitary Error Bases and
Quantum Codes,'' {\em LANL Preprint}, quant-ph/9608048, 1996. 

\bibitem{Knill96_2} E. Knill, ``Group Representations, Error Bases and
Quantum Codes,'' {\em LANL Preprint}, quant-ph/9608049, 1996. 

\bibitem{knill97}
E. Knill and R. Laflamme, ``A theory of quantum error correcting codes,''
{\em Phys. Rev. A}, vol. 55, pp. 900-911,
   1997.

\bibitem{knill99}
E. Knill, R. Laflamme and L. Viola, ``Theory of quantum error correction
for general noise'', \emph{Phys. Rev. Lett.}, vol. 84, pp. 2525-2528, 2000.

\bibitem{macw77}
F.\,J.\,MacWilliams and N.\,J.\,A.\,Sloane,
{\em The Theory of Error-Correcting Codes},
New York: North-Holland, 1977.

\bibitem{rains} E. Rains, Nonbinary quantum codes, LANL e-print
quant-ph/9703048.

\bibitem{shor94} P.W.~Shor, ``Polynomial-time algorithms for prime
factorization and discrete logarithms on a quantum computer,''
{\em Proceedings of the 35th Annual Symposium on the Foundations of Computer Science},
S.Goldwasser, Editor, IEEE Computer Society Press, Los Alamitos, CA, p.124,
1994.

\bibitem{shor95} P.W.~Shor, ``Scheme for reducing decoherence in quantum
memory,'' Phys. Rev. A,
{\bf 52}, p. 2493, 1995.

\bibitem{ref 1}
P.W.~Shor and R. Laflamme, ``Quantum analog of the MacWilliams identities
in classical coding theory,'' {\em Phys. Rev. Lett.}, vol. 78, pp. 1600-1602, 
1997.

\bibitem{stean96_1}    
A. M. Steane,  "Simple quantum error correcting codes," {\em Phys. Rev. Lett.},vol. 77, pp. 793-797, 1996.

\bibitem{stean96_2}
A. M. Steane, "Multiple particle interference and quantum
   error correction," {\em Proc. Roy. Soc. London A}, vol. 452, pp.
   2551-2577, 1996.





\end{thebibliography}
\end{document}